\documentstyle[preprint,tighten,aps,epsfig]{revtex}

\def\lapprox{\mathrel{\mathop  {\hbox{\lower0.5ex\hbox{$\sim$}
\kern-0.8em\lower-0.7ex\hbox{$<$}}}}}  
\def\gapprox{\mathrel{\mathop  {\hbox{\lower0.5ex\hbox{$\sim$}
\kern-0.8em\lower-0.7ex\hbox{$>$}}}}}

\begin{document}

\draft

\preprint{\vbox{\noindent{}\hfill INFNFE-20-01}}

\title{What have we learnt about the Sun from the measurement 
of the $^8$B neutrino flux?}

\author{G.~Fiorentini$^{(1,2)}$ and B.~Ricci$^{(1,2)}$}

\address{
$^{(1)}$Dipartimento di Fisica dell'Universit\`a di Ferrara,I-44100
Ferrara,\\
$^{(2)}$Istituto Nazionale di Fisica Nucleare, Sezione di Ferrara, 
I-44100 Ferrara,}

\maketitle

\begin{abstract}

By combining the results of SNO and Super-Kamiokande one can
 derive - in the absence of sterile neutrinos - the total neutrino
 flux produced from $^8$B decay in the Sun. We use this information
 to check the accuracy of several input parameters of solar 
model calculations. Opacity and p-p fusion cross section
 are constrained by the $^8$B flux measurement to
 the level of few per cent. The central solar temperature 
is determined to the one-percent level.
 We also find an upper limit for the flux on Earth of
 sterile neutrinos. We discuss the role of nuclear physics 
uncertainties on these determinations.

\end{abstract}

\section{Introduction}
\label{intro}

Electron neutrinos from  $^8$B decay in the Sun have
 been detected at the Sudbury Neutrino Observatory (SNO)
 by means of the charged current (CC) reaction on deuterium \cite{sno}. 
The result, $\Phi_e=(1.75\pm 0.14)$ 10$^6$ cm$^{-2}$ s$^{-1}$ 
(here and in the following one sigma statistical errors 
are combined in quadrature with systematical errors), 
is a factor three smaller than the SSM prediction of Bahcall et al.\cite{bp2000}:
\begin{equation}
\label{fibssm}
\Phi^{SSM}= 5.05 \cdot (1^{+0.20}_{-0.16})\cdot 10^6 \,{\rm cm}^{-2}\,{\rm s}^{-1} \quad .
\end{equation}

The CC reaction is sensitive exclusively to $\nu_e$, while 
Electron Scattering (ES) also has a small sensitivity 
to $\nu_\mu$ and $\nu_\tau$. Comparison of $\Phi_e$ to the Super-Kamiokande (SK)
 precision result on ES yields a 3.3$\sigma$ difference, 
providing evidence that there is a non-electron flavor 
active neutrino component in the solar flux. 

Extraction of this flux, $\Phi_{\mu+\tau}$, can be done in a 
model independent way by exploiting the similarities of 
the response functions of SNO and of SK, see \cite{VFL,nove}.
In this way, one determines from the two experiments the total 
active neutrino flux, $\Phi^{EXP}= \Phi_e+\Phi_{\mu+\tau}$, 
produced by $^8$B decay in the sun \cite{Fogli}:
\begin{equation}
\label{fibfogli}
	\Phi^{EXP} = 5.20 \cdot (1^{+0.20}_{-0.16})\cdot 10^6 \,{\rm cm}^{-2}\,{\rm s}^{-1} \quad .
\end{equation}

The close agreement with the theoretical prediction (\ref{fibssm}) is
 an important confirmation of the robustness of SSM calculations, 
see Table \ref{tabtc}.

Actually, one can go somehow in more detail and use this 
experimental information to provide an independent estimate 
of the accuracy of several input parameters that are used in 
solar model calculations.  Among others, we shall consider the
 solar metal abundance, opacity and p-p fusion cross section.

We shall also consider the central solar temperature.
 As well known, this is not an independent parameter, 
rather it is the result of solar model calculations,
 mainly dependent on the assumed values for metals, 
opacity and luminosity. The boron flux measurement 
essentially provides a measurement of the central solar temperature.

So far we neglected the possibility of oscillations 
of $\nu_e$ into sterile neutrinos. In fact one can use 
the experimental result on active neutrinos $\Phi^{EXP}$
 and the theoretical prediction on the total flux, 
$\Phi^{SSM}$ to provide an upper bound on the flux of 
sterile neutrinos on earth.
 
When connecting the $^8$B flux with quantities 
characterizing the solar interior, an important
 link is provided by nuclear physics, through 
the nuclear cross sections of the processes 
which lead to $^8$B production.  We shall discuss
 the relevance of nuclear physics uncertainties in this respect.

\section{Power laws}
\label{power}

The neutrino flux from $^8$B decay in the Sun depends on several
 inputs $Q_i$, which can be grouped as nuclear and astrophysical
 parameters, see Table \ref{tab1}.

The first group contains the zero energy astrophysical
 S-factors for the reactions which are involved in
 the production of $^8$B nuclei. $S_{11}$ refers to
 p+p$\rightarrow$d+ e$^+$ +$\nu$,
 $S_{33}$ and $S_{34}$ refer respectively to 
$^3$He+$^3$He $\rightarrow ^4$He+2p and to
$^3$He+$^4$He $\rightarrow$ $^7$Be+$\gamma$.
  Once $^7$Be nuclei have been formed,
 the probability of forming a $^8$B nucleus is essentially
 given by the ratio of proton to electron capture on $^7$Be, i.e on $S_{17}/S_{e7}$.

The $^8$B neutrino flux depends on several astrophysical inputs. 
Concerning
 the dependence on the solar luminosity 
$L_\odot$ and  age $t_\odot$,
 a more luminous Sun requires a higher internal temperature, 
which implies a higher Boron flux. An older Sun is also 
a more luminous sun.  

The metal content Z/X determines opacity. 
 If Z/X increases, opacity increases  and the radiative 
transfer of the solar luminosity requires higher temperature 
gradients, which in turn implies higher temperature and a larger $^8$B flux.

Opacity calculation are complex, and also for fixed metal 
content different codes yield (slightly) different opacities. 
For this reason we introduce, by means of a scaling parameter $\kappa$, 
the possibility of a uniform scaling of opacity in the solar interior.

  Another scaling factor, D, is allowed for varying the 
calculated diffusion coefficients.

If these inputs are changed with respect to the values used in
 the SSM calculations, $Q_i^{SSM}$, the $^8$B-neutrino flux 
$\Phi$ changes according to a power law:
\begin{equation}
\label{eqpower}
 \Phi =\Phi^{SSM} \Pi  (Q_i/Q_i^{SSM} )^{\alpha_i} \quad .
\end{equation}

We have calculated these coefficients by using solar models
which include helium and heavy elements diffusion, see Table \ref{tab1}.
 They are in  agreement with the values calculated (neglecting diffusion)
 in  \cite{libro} (see Table \ref{tab1}) and in \cite{report}.

Most of these values can be understood by using simple, 
semi-quantitative arguments, see e.g. \cite{report}.
 As an example, one can easily show that at equilibrium
 the production of $^7$Be nuclei scales as
 $S_{34}/\sqrt{S_{33}}$   \cite{scilla,report} and that the production of $^8$B nuclei,
 which is a minor perturbation of the pp-II chain, scales as $S_{17}/S_{e7}$.

Each input is affected by uncertainties. The values shown in Table \ref{tab1},
 4-th row,  correspond to one sigma percentage error. 
They have been obtained according to the following criteria:

i) For the nuclear cross sections, we generally consider the
 values and uncertainties recommended in ref. \cite{adelberger}
 which are quite similar to those quoted in ref. \cite{nacre}. 
Concerning $S_{17}$, we quote an uncertainty of 9\%, taking 
into account recent measurements \cite{s17}.

ii) Concerning solar luminosity, age and metals, we adopt 
the same uncertainties quoted in \cite{bp95}, which are derived by 
recent measurements of the solar constant, by a detailed
 comparison of meteorite radioactive datings and by analysis
 of meteoric and photospheric chemical compositions \cite{grevesse}.

iii) Regarding the opacity and diffusion uncertainties, there
 is no experimental guidance and one has to resort to the
 comparison among different theoretical calculations. 
A factor 2.5\% is the typical difference between the 
calculated values of the Livermore and Los Alamos opacities, see \cite{bp92}.
 Concerning diffusion, Bahcall and Loeb \cite{loeb} quote 
and uncertainty of about 15\% by comparing results 
of different theoretical calculations.
(We have found that agreement with helioseismic information fixes
diffusion to the level of 10\% \cite{diffusion}.)

We also show in Table \ref{tab1}, fifth line, the contributed fractional 
error, $ \alpha \, \Delta Q/Q$, i.e the contribution of each parameter 
to the error on  the neutrino flux.
Summing up in quadrature, the nuclear uncertainties yield an
 error on the flux $(\Delta \Phi/ \Phi)_{nuc}$=13.3\%, 
which matches
 closely the error resulting from astrophysical uncertainties
 calculated according to the same prescription,  $(\Delta \Phi/ \Phi)_{ast}$=12.2\%.
All in all, the total calculated 
uncertainty,
$(\Delta \Phi/ \Phi)= \sqrt{(\Delta \Phi/ \Phi)_{nuc}^2  + (\Delta \Phi/ \Phi)_{ast}^2}$
 is about 18\%, in agreement with ref. \cite{bp2000}.

\section{Results}
\label{result}

The experimental result, Eq. (\ref{fibfogli}), is in excellent agreement with
 the theoretical prediction, Eq. (\ref{fibssm}). The two determinations are
 affected by a similar error and, as we have just seen, the 
theoretical error gets comparable contributions form nuclear 
and astrophysical uncertainties.

One can expect that in the future, with increasing statistics 
and with better understanding of the systematical uncertainties,
the experimental error will be reduced, possibly by a factor of two.
In view of this, one should improve the theoretical 
calculation by a comparable factor, which requires a better
 understanding of the nuclear physics and astrophysics which
 are involved, see again Table \ref{tab1}.

We remind that experiments are only sensitive to active neutrinos.
 The comparison with the SSM prediction can be used to derive an 
upper bound on the presence of sterile neutrinos, by
 requiring that $\Phi_{s}= \Phi^{SSM}- \Phi^{EXP}$. This gives:
\begin{equation}
\label{sterile}
\Phi_{s} < 2.5\cdot 10^6 \, {\rm cm}^{-2} \, {\rm s}^{-1} 
\qquad ({\rm at}\, 2\sigma ) \qquad .
\end{equation}

Improvement of this bound will require that both theoretical
 and experimental determinations become more accurate.

Neglecting the possibility of sterile neutrinos, one can use 
the $^8$B-flux measurement as an independent way of estimating 
the accuracy of several nuclear and astrophysical parameters.

In fact, the measurement of $\Phi$ can be interpreted as a way of 
determining each of the parameters listed in Table \ref{tab1}, by means of:
\begin{equation}
\label{qi1}
 Q_i = \left ( \frac {\Phi^{EXP}}{\Phi^{SSM}} \right ) ^{1/\alpha_i} 
\Pi_{j \neq i}\left ( \frac{Q_j}{Q_j^{SSM}} \right ) ^{-\frac{\alpha _j}{\alpha _i}}
\end{equation}
In this way  the measurement  of $\Phi$ together with  the 
independent  information on $Q_j$ ($j \neq i$) can be translated  into 
a mesurement of $Q_i$.

For each parameter in Table \ref{tab1}, we have estimated the 
corresponding accuracy by taking into account the $^8$B
 result and the uncertainty on all other parameters, 
in turn.  The resulting values,
\begin{equation}
\label{deltaqi}
 (\Delta Q_i/Q_i)_B = \frac{1}{\alpha _i}
\sqrt{
\left (\frac{ \Delta \Phi}{\Phi} \right ) ^2 + 
 \Sigma_{j \neq i} \left ( \alpha_j \frac{\Delta Q_j}{Q_j} \right )^2 } \quad ,
\end{equation}

are listed in Table \ref{tab1}, last row . 
We remark the following points:
\begin{itemize}

\item 
The opacity scaling factor $\kappa$ is determined  to be unity 
with an accuracy of  9.3\%. This is somehow worse than the estimated
input accuracy of 2.6\%. We remind however that this uncertainty was
essentially estimated form the comparison between two different
theoretical calculations.

\item 
The metal content  Z/X is confirmed within 18\%. Again this is
worse than the estimated input uncertainty (6.1\%) from studies of solar
photosphere, which however might not be representative of the metal
content of the solar interior if it has been enriched in metals  after
the sun formation.
  
\item
 The astrophysical $S$-factor for the pp fusion reaction, $S_{11}$, is
determined wih accuracy of 9\%. This quantity is not
measured and it is the result of theoretical errors. The  estimated
uncertainty of these calculations (``input error") is  1.7\%
Helioseismology allows a few per cent  accuracy \cite{noispp}.

\item 
The errors on $S_{33}$, $S_{34}$, $S_{e7}$,
 $S_{17}$ , $L_\odot$ and $t_\odot$ are not
competitive with more direct measurements in the literature.

\end{itemize}

\section{The central solar temperature}
\label{sectemp}

A final remark concerns the accuracy of the central solar 
temperature $T_c$ (see Table \ref{tabtc} 
for theoretical predictions of $T_c$ ).

 Solar model builders have been claiming for decades 
that $T_c$ is known with an accuracy of one per cent or better.
Helioseismology has provided strong support to this finding. Although
helioseismology measures sound speed and not temperature, consistency
with helioseismic data has been found only for solar models which
produce $T_c$ within one percent of the SSM prediction \cite{hcsm,bahctc}.
The measurement of $^8$B provides an important test in this
respect.

In fact, the temperature is not an independent variable. 
Its precise value is determined by some of the parameters 
which we have been discussing: the cross section for the 
pp-reaction, the metal content of the Sun, the adopted 
values for the radiative opacity, the solar age and 
luminosity and also the diffusion coefficients. 

 On the other hand, the central temperature
 is not affected by the other nuclear parameters listed in Table \ref{tab1}.
 In full generality, the relationship between the boron flux and
  $T_c$ is (see \cite{report}):
\begin{equation}
\label{fitc}
 \Phi = \Phi^{SSM} \left ( \frac{T_c}{T_c^{SSM}} \right )^{\beta} 
\frac{S_{nuc}} {S_{nuc}^{SSM} } \quad ,
\end{equation}
Where  $S_{nuc}$ includes the dependence on the nuclear cross sections

\begin{equation}
\label{snuc}
	S_{nuc} = S_{34}^{0.84} S_{33}^{-0.43} S_{17}/ S_{e7}
\end{equation}

As discussed above, $S_{nuc}$ is determined with an accuracy 
of 13.3\%. 

\begin{figure}
\caption[ff]{The behaviour of the $^8$B neutrino flux as a function of the 
central temperature, when varying different solar model inputs of $\pm 10\%$.
The straight line corresponds to  the power law $T_c^{20}$.}
\epsfig{figure=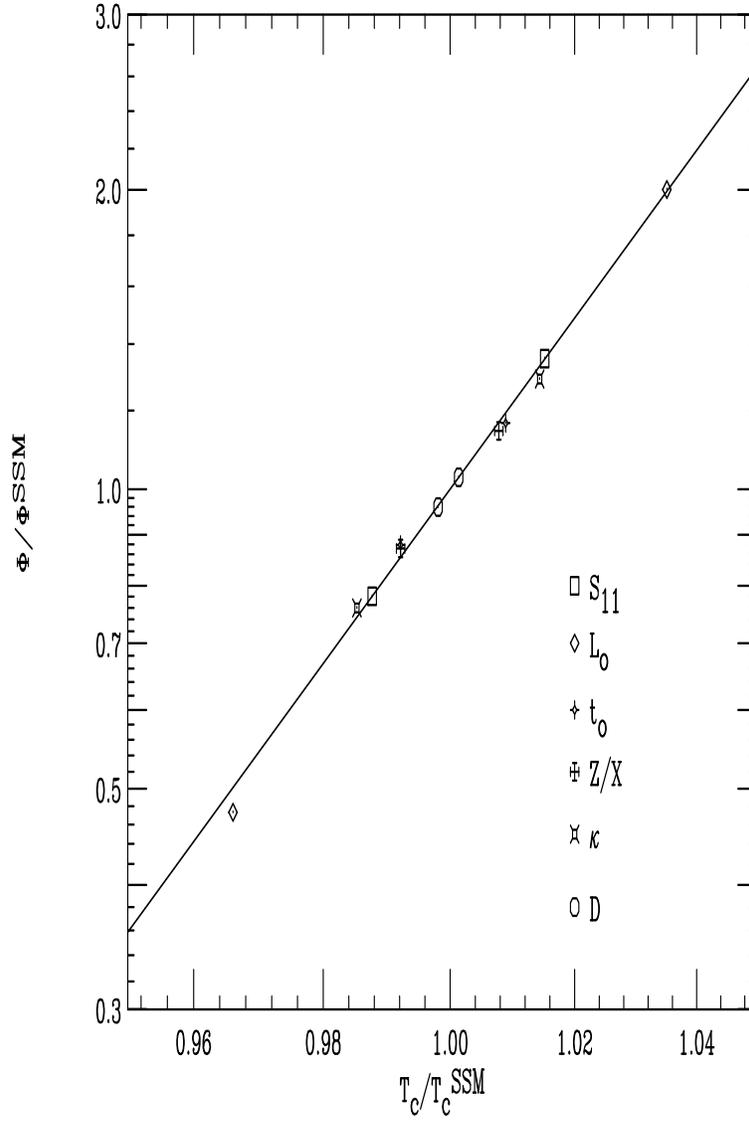,height=10cm,width=15cm,angle=90}
\label{figbeta}
\end{figure}

Concerning the exponent $\beta$ in Eq. (\ref{fitc}), it is weakly dependent
on which parameter is being varied  for obtaining a changement of $T_c$.
We have calculated the exponent  by using solar models which include
 helium and heavy elements diffusion,
see Table \ref{tabeta} and Fig. \ref{figbeta}.
On these grounds we will take $\beta=20$. We remark that if one
neglects diffusion, slightly higher values of $\beta$ are obtained,
see  Table \ref{tabeta} last column. In fact in ref. \cite{bu96}
by studying  solar models without diffusion  a somehow higher value
was found, $\beta=24$. 
Anyhow the choice of a slightly different value
dos not induce a significant difference in the conclusions.

The agreement between theory and experiment on
 $\Phi$ thus implies that the central temperature of the Sun 
agrees with the SSM prediction to within one per cent:
\begin{equation}
\label{tc}
			T_c =15.7 (1\pm 1\%) 10^6 K \quad ,
\end{equation}
where the error gets comparable contributions from 
the uncertainty on $\Phi$ and from nuclear physics.

\acknowledgments

G.F. is grateful to the CERN theory division for hospitality.
We are grateful to John Bahcall for useful comments and suggestions.

\begin{table}
\caption[bb]{Predictions of some SSM calculations:}
\begin{tabular}{c|c|c|c|c|c}
&  BP2000 \cite{bp2000} & FRANEC97 \cite{FRANEC97} & RCVD96 \cite{RCVD96} 
& JCD96 \cite{sunjcd}& GARSOM2 \cite{Schlattl}\\ 
\hline
$T_c \; [10^6 \rm{K}]$ &  15.696   &  15.69 & 15.67 & 15.668 & 15.7\\  
$\Phi_{\rm B} \; [10^{6}{\rm cm}^{-2} {\rm s}^{-1}]$
                       &  5.05  &  5.16 &   6.33 & 5.87 &  5.30 \\
%$\Phi_{\rm Be} \; [10^{9}{\rm cm}^{-2} {\rm s}^{-1}]$
%                       &  4.77  &  4.49  &   4.8  & 4.94 &  4.93 \\
\end{tabular}
\label{tabtc}
\end{table}

\begin{table}
\caption[aa]{Nuclear and Astrophysical parameters related to the determination  of $^8$B flux}
\begin{tabular}{l|ccccc|ccccc}
       &\multicolumn{5}{c|}{Nuclear} & \multicolumn{5}{c}{Astrophysical} \\
$Q$ & $S_{11}$  &  $S_{33}$  &  $S_{34}$  &  $S_{e7}$  &  $S_{17}$  &
      $L_{\odot}$  & $t_{\odot}$ & Z/X  &  $\kappa$   &  D \\
\hline
$\alpha$              & -2.7  & -0.43 & 0.84 & -1 & 1 &  7.2 & 1.4  & 1.4  & 2.6 & 0.34 \\
$\alpha$ \cite{libro} & -2.59 & -0.40 & 0.81 & -1 & 1 & 6.76 & 1.28 & 1.26 &  -   & -    \\
Input error (\%)      & 1.7 & 6.1 & 9.4 & 2 & 9 & 0.4 & 0.4 & 6.1 & 2.5 & 15 \\
Contributed error to $\Phi$  (\%) & 4.6 & 2.6 & 7.9 & 2 & 9 & 2.9  & 0.56 & 8.5 & 6.5 & 5 \\
\hline
{\bf Uncertainty derived from}   & & & & & & & & & &\\
{\bf $^8$B 
measurement (\%)}     & {\bf 9 } & {\bf 58 }  & {\bf 28 } & 
{\bf 25 } & {\bf 23 } & {\bf 3.5 } & {\bf 18 } & {\bf 18 } & {\bf 9.3 } &{\bf 73 }  \\
\end{tabular}
\label{tab1}
\end{table}

\begin{table}
\caption[cc]{Dependence of the boron flux from central temperature, 
see Eq. (\ref{fitc})}
\begin{tabular}{lll}
$Q_i$  		&  $\beta$ (with diffusion) & $\beta$ (without diffusion) \cite{report}\\
\hline
$S_{11}$  	&  19.5   & 21\\
$L_\odot$	&  21.1   &   \\
$t_\odot$	&  17.0   & 18\\
$Z/X$		&  17.4   & 21\\
$\kappa$        &  18.8   & 19\\
D		&  21.2   &   \\
\end{tabular}
\label{tabeta}
\end{table}

\end{document}